# Tuning the hysteresis loop for the anomalous Hall effect in Pt ultrathin films on CoFe$_2$O$_4$ by electrolyte gating

Leon Sakakibara[1], Yumiko Katayama[1], Tomohiro Koyama[2,3,4], Daichi Chiba[2,3,4], and Kazunori Ueno[1]

[1] Graduate School of Arts and Sciences, University of Tokyo, Meguro, Tokyo 153-8902, Japan

[2] Institute of Scientific and Industrial Research, Osaka University, Ibaraki, Osaka 567-0047, Japan

[3] Center for Spintronics Research Network, Osaka University, Toyonaka, Osaka 560-8531, Japan

[4] Spintronics Research Network Division, Institute for Open and Transdisciplinary Research Initiatives, Osaka University, Suita, Osaka 565-0871, Japan

**Abstract**

Pt ultrathin films on ferromagnetic insulators have been widely studied for spintronics applications, and magnetic moments of interface Pt atoms were considered to be ferromagnetically ordered due to a magnetic proximity effect (MPE). An anomalous Hall effect (AHE) is usually used to examine an out-of-plane magnetic moments of the Pt layer. To tune ferromagnetic properties of an Pt ultrathin film, we fabricated electric double layer transistors on Pt thin films with thicknesses of 5.9 nm and 7.0 nm on a CoFe$_2$O$_4$ (CFO) ferrimagnetic insulator. For the Pt (7.0 nm)/CFO sample, a hysteresis loop was observed in the anomalous Hall resistivity without the gate bias, and the coercive field was tuned by applying the gate bias. For the Pt (5.9 nm)/CFO sample, a hysteresis loop was not observed without a gate bias, but was opened by applying a gate bias ($V_G$ = ±3 V). This indicated that the long-range ferromagnetic ordering of magnetic moments in the Pt film was switched on and off by the electric field effect. The hysteresis loop was observed up to 19.5 K for a $V_G$ of +3 V, while the AHE was observed up to approximately room temperature.

Pt has a density of states $D(E)$ peak near the Fermi energy $E_F$ and has been considered to be close to the Stoner criteria, which is fulfilled by the ferromagnetic metals Fe, Co, Ni [1]. There have been many reports on anomalous Hall effects (AHEs) in Pt/ferromagnetic insulator (FMI) structures [2] [3] [4] [5] [6] [7] [8], which have been attributed to out-of-plane magnetic moments of Pt atoms due to the magnetic proximity effect (MPE). The Pt/FMI interface has also been widely studied as a model system for examining the spin Hall effect (SHE) [9] [10] [11] [12] [13], which also contributes anomalous Hall resistance in the Pt layer. A contribution from the SHE was distinguished from that of the AHE with the magnetic moments in Pt, since ferromagnetic properties, such as the Curie temperature $T_C$ and a coercive field $B_c$, of the FMI layer were different from those in the Pt layer. In addition, magnetic moments in Pt atoms close to the FMI interface have been confirmed in various ways, such as with X-ray magnetic circular dichroism data [14] and theoretical calculations [7].

The electric field effect has been used to induce ferromagnetism and to tune ferromagnetic properties of metallic thin films [5] [15] [16] [17] [18] [19] [20] [21] [22]. S. Nodo, *et al.* reported that the magnitude of the anomalous Hall resistance of a Pt (2.5 nm)/spinel $CoFe_2O_4$ (CFO) bilayer structure was changed by -50% and +100% with gate biases of +2 V and -2 V, respectively, while $B_c$ remained constant [21]. K. T. Yamada, *et al.* reported that both a Fermi level and a magnetic moment of Pt indeed changed with gating on Pt (0.4 nm)/Co structure using XMCD and XAS [22]. In addition, there are several reports on electric field-effect tuning of the anomalous Hall effect in Pt/ $Y_3Fe_5O_{12}$ (YIG), while Pt/YIG always showed zero spontaneous magnetization with and without the gate bias. However, there have been no reports on tuning $B_c$ or $T_C$ or inducing spontaneous magnetization with the electric field effect on Pt/FMI structures, although this is strongly desired for spintronics applications.

In this study, Pt/CFO samples with various Pt thicknesses were prepared. A pristine Pt (5.9 nm)/CFO sample had no hysteresis loop in the anomalous Hall resistance at 2 K, while a pristine Pt (7.0 nm)/CFO sample showed a hysteresis loop at 2 K. Then, electric double layer transistors (EDLT) were fabricated on these films. By using the ionic liquid (IL) DEME-TFSI, high-density charge carriers accumulated on the films, and the Fermi level on a top layer of the film was tuned. Then, the charge accumulation opened a hysteresis loop in the anomalous Hall resistance for the Pt (5.9 nm)/CFO sample at 2 K. The temperature dependence and gate bias dependence of the ferromagnetic properties were also studied with these samples. In addition, a Pt (5.9 nm) sample on a nonferromagnetic $SrTiO_3$ (STO) substrate was also studied for comparison.

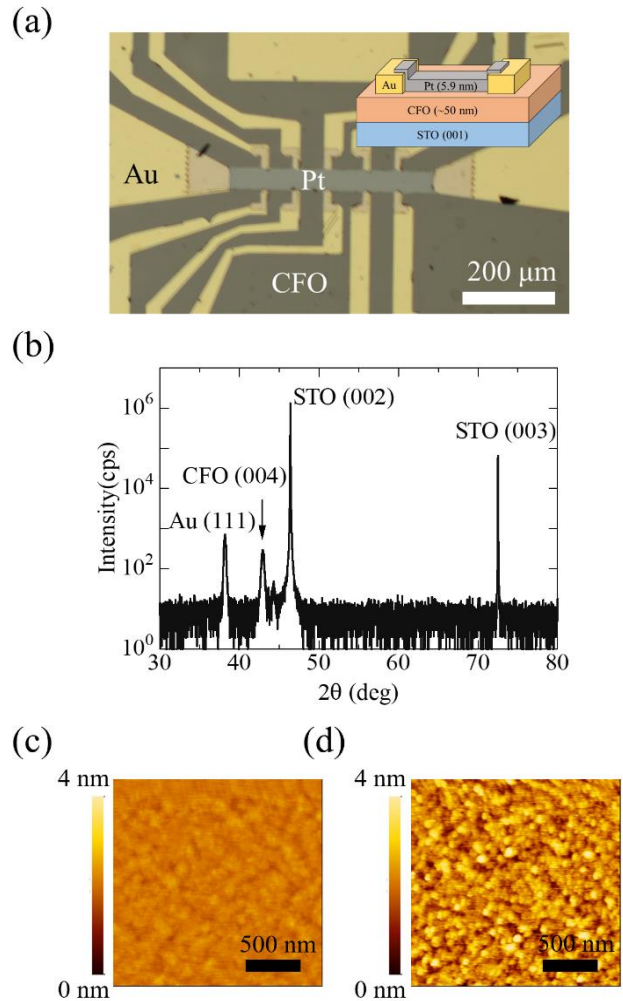

Fig. 1. (a) Schematic illustration of the Pt (5.9 nm)/CFO sample with Au electrodes. (b) XRD pattern of the Pt (5.9 nm)/CFO sample with Au electrodes. (c), (d) AFM images of the CFO film and Pt (5.9 nm)/CFO samples, respectively.

Schematic structures of the samples are shown in the inset of Fig. 1(a). CFO films with a thickness of 50 nm were grown on STO (001) substrates by pulsed laser deposition using a KrF excimer laser. During deposition, the oxygen pressure $P_{O_2}$ was maintained at $10^{-4}$ Torr, and the temperature of the substrates was maintained at 800 °C. After deposition, the substrates were annealed at 750 °C and under a $P_{O_2}$ of $10^{-2}$ Torr for 30 min. Contact electrodes were prepared on the CFO film with an electron-beam evaporated Au (100 nm)/Ti (20 nm) film. Then, a Pt film with a thickness of 5.9 nm or 7.0 nm was deposited by sputtering through a Hall-bar-shaped photoresist mask. The sample without the ionic liquid was called the "pristine" sample, as shown in Fig. 1(a). Finally, the IL, [DEME]$^+$[TFSI]$^-$, was placed directly on the channel area of the Hall bar. The Pt (5.9 nm)/STO sample was also prepared in a similar way. The crystallinity and surface structure were confirmed by X-ray diffraction (Smartlab, Rigaku) and atomic force microscopy (AFM), respectively. Film thickness was confirmed by a surface profilometer (DekTak, Bruker). Magnetic and transport properties were examined with a superconducting quantum interference device (SQUID) and a Physical Property Measurement System (PPMS, Quantum Design), respectively.

Figure 1(b) shows the $\theta - 2\theta$ X-ray diffraction (XRD) pattern of the Pt (5.9 nm)/CFO sample. The peaks at $2\theta = 46.4°$ and $72.5°$ were attributed to STO (001), and the peak at $2\theta = 38.2°$ was attributed to the Au electrode. The peak at $2\theta = 42.9°$ was for CFO (004), from which the estimated out-of-plane lattice constant is 8.417 Å. This confirmed an out-of-plane tensile strain of 0.30% for the film compared to the bulk value (8.392 Å [23]), which is consistent with previous reports [24]. Additionally, no peaks were identified for the Pt film, suggesting that the Pt film was polycrystalline. Fig. 1(c) and 1(d) show AFM images (2 μm × 2 μm) of the CFO and Pt films. The RMS roughnesses were confirmed to be 0.139 nm and 0.630 nm, respectively. Since the roughness of both the Pt/CFO interface and the Pt film surface was much smaller than the Pt thickness, the Pt was confirmed to form a continuous film. Magnetic field dependence of magnetization at 300 K (see Fig. S1 in the Supplementary Information) showed that the CFO film was a ferromagnet at 300 K and had in-plane anisotropy with $B_c$ of 320 mT. With a perpendicular magnetic field, $B_c$ was approximately 50 mT at both 300 K and 2 K.

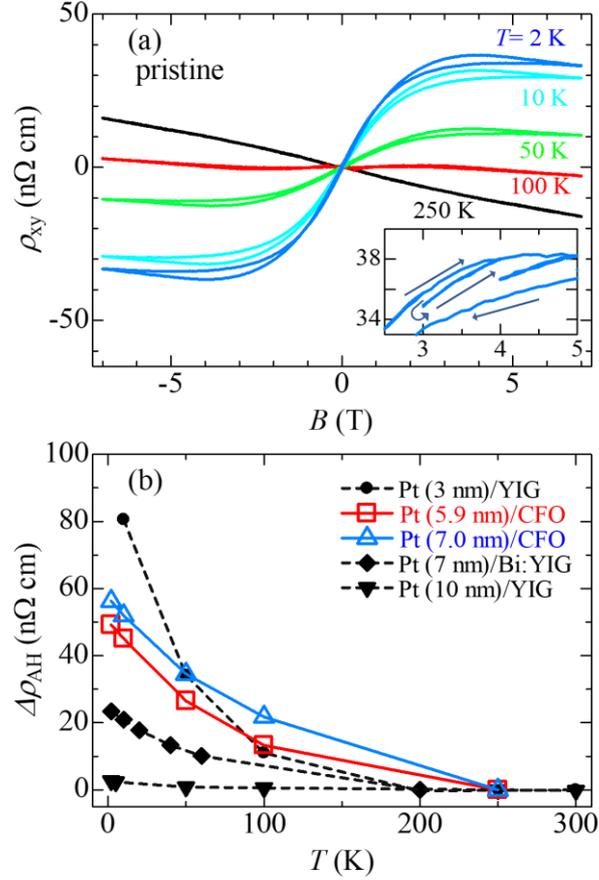

Fig. 2. (a) Magnetic field dependence of Hall resistivity $\rho_{xy}$ for the pristine Pt (5.9 nm)/CFO sample at various temperatures. The inset shows a hysteresis loop in $\rho_{xy}$ at a high magnetic field with changing $B$ as 0 T→4 T→3 T→5 T→4 T→7 T→0 T at 2 K. (b) Difference in $\rho_{AH}$ at 7 T, $\Delta\rho_{AH}$, as a function of temperature. Data from previous studies, Pt (10 nm)/YIG, [25], Pt (7 nm)/Bi:YIG [5], and Pt (3 nm)/YIG [8], were also plotted. $\Delta\rho_{AH}$ was defined as the difference from the value at a reference temperature. The reference temperatures were 250 K, 200 K, 250 K, and 300 K for this study, reference [5], reference [25] and reference [8], respectively.

We examined the magnetic field dependence of the Hall resistance for the pristine Pt (5.9 nm)/CFO sample and observed the AHE at all temperatures. The Hall resistance $\rho_{xy}$ is expressed as the sum of the ordinary Hall term $\rho_{OH} = R_H B$, which is proportional to the magnetic field $B$, and the anomalous Hall term $\rho_{AH}$ as follows:

$$\rho_{xy} = \rho_{OH} + \rho_{AH} \qquad (1)$$

Figure 2(a) shows the magnetic field dependence of $\rho_{xy}$ at various temperatures for the pristine Pt (5.9 nm)/CFO sample. We observed AHE without any hysteresis loop at 0 T from 250 K to 2 K, and the absolute value of $\rho_{AH}$ increased with decreasing temperature. In addition, the sign of $\rho_{AH}$ was inverted from 250 K to 100 K. This sign reversal from negative at approximately room temperature to positive at low temperature has generally been reported for Pt/YIG systems [5] [6] [8] [25] and will be discussed later. Figure 2(b) shows the temperature dependence of the difference in $\rho_{AH}$ at 7 T, $\Delta\rho_{AH}$, from the value at approximately room temperature. Data from previous studies on Pt (10 nm)/YIG [25], Pt (7 nm)/$Bi_{0.6}Y_{2.4}Fe_5O_{12}$ (Bi:YIG) [5], and Pt (3 nm)/YIG [8] are also shown. All data showed increases in $\Delta\rho_{AH}$ with decreasing temperature, and larger changes were systematically observed for thinner Pt films.

A change in the sign of the anomalous Hall resistance with temperature change has been observed in many previous studies on Pt/YIG systems [5] [6] [8] [25]. In a study of Pt/$Y_3Fe_{5-x}Al_xO_{12}$ [6], the sign of $\rho_{AH}$ was determined by a competition between MPE ($\rho_{AH} \propto M_z(Pt)$) and SHE ($\rho_{AH} \propto -M_z(YIG)$), where $M_z$ is the out-of-plane component of the magnetization. The sign reversal occurred because the MPE and SHE were dominant at low and high temperatures, respectively. In a study of the electric field effect on a Pt film, a similar AHE inversion was observed for a Pt thin film without a FMI layer [5]. The authors explained that both negative and positive components existed in $\rho_{AH}$, and the sign reversal occurred due to the competition between the two components. Both studies agreed that a negative $\rho_{AH}$ at low temperature originated from out-of-plane magnetic moments in Pt, which were induced by MPE or field effects. We thus assume that $\rho_{AH}$ below 100 K originates from the out-of-plane magnetic moment in Pt in the following text.

It is noted that hysteresis behavior in $\rho_{xy}$ was observed for magnetic fields above 1 T below 100 K. The inset of Fig. 2(a) shows details of the hysteresis at 2 K. When the magnetic field was increased and then decreased, $\rho_{xy}$ showed a clockwise (CW) hysteresis curve. Then, when the magnetic field returned to the original value, $\rho_{xy}$ returned to the original value along the same curve as the magnetic field was decreased. In ferromagnets, magnetization always shows a counterclockwise (CCW) hysteresis since magnetization is induced by increasing the magnetic field and remains with decreasing magnetic field. Other magnetic materials (e.g., metamagnetic materials) may also exhibit hysteresis in magnetization, but the direction of the hysteresis is always CCW. Thus, such ferromagnetic magnetization cannot explain the CW hysteresis in our sample. We have two explanations for this CW hysteresis. One is an anomalous Hall resistance originating from the SHE. The $\rho_{AH}$ in Pt is proportional to the out-of-plane magnetization -$M_z$ of CFO, and CW hysteresis occurs in $\rho_{AH}$. However, since the saturation magnetic field of the CFO was below 1 T, the CW hysteresis above 1 T cannot be explained by magnetization in the CFO. Another explanation is complex magnetic domains in a nanostructured ferromagnet. Similar CW hysteresis has been reported in ferromagnetic nanoparticles that connect each other with ferromagnetic or antiferromagnetic interactions. [26] [27]. Our Pt thin films may have nonuniform thicknesses and result in a mixture of nanoscale ferromagnetic and paramagnetic sites. This nanostructure could result in complex magnetic domains similar to those of nanoparticles, and CW-hysteresis may occur.

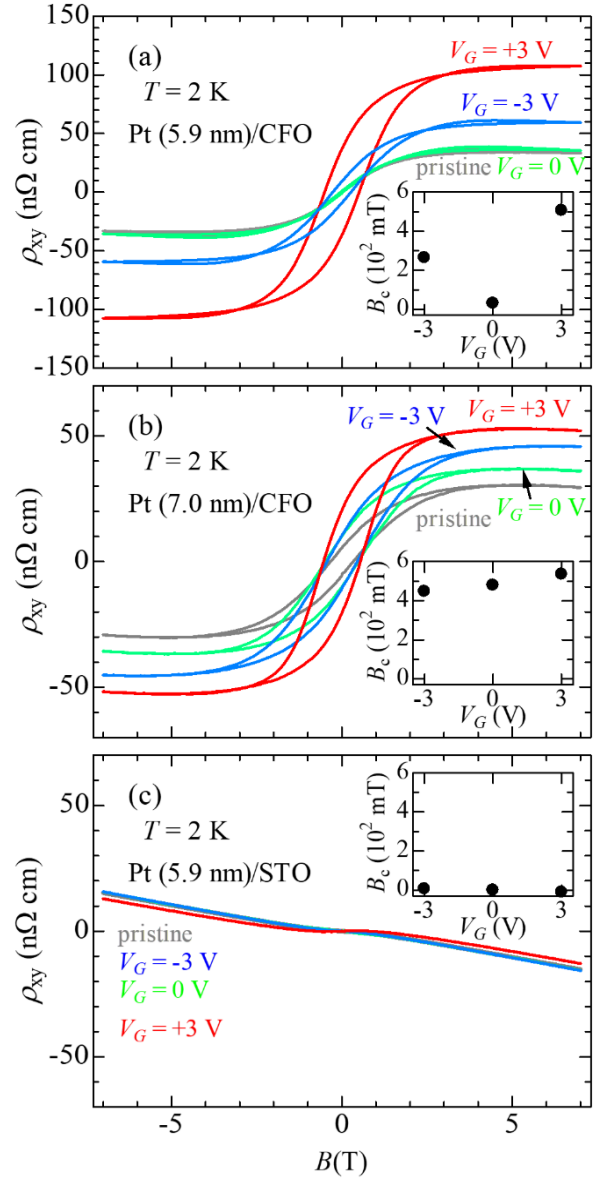

Fig. 3. Magnetic field dependence of $\rho_{xy}$ for various gate biases $V_G$ at 2 K for (a) Pt (5.9 nm)/CFO, (b) Pt (7.0 nm)/CFO, and (c) Pt (5.9 nm)/STO samples. Gray, green, blue, and red lines correspond to data for pristine, $V_G = 0$ V, -3 V, and +3 V, respectively. The inset shows a coercive field $B_c$ as a function of $V_G$. For the Pt (5.9 nm)/CFO sample, $V_G$s were applied in the order of 0 V, +3 V, -3 V, +3 V. The first and second applications of +3 V reproduced $\rho_{xx}$. Detailed measurements were made during the second application, which are shown in the figure. For the other samples, measurements were made in the order of 0 V, +3 V, -3 V.

The $V_G$ dependence of transport properties was examined for various samples. By applying a positive and negative $V_G$, the four-terminal conductivity increased and decreased, respectively, corresponding to n-type FET behavior and indicating charge accumulation on the Pt film (see Fig. S2 in the Supplementary Information). Figure 3(a) shows the magnetic field $B$ dependence of $\rho_{xy}$ measured for a sample fabricated on the pristine Pt (5.9 nm)/CFO sample. The pristine sample and the sample with a $V_G$ of 0 V showed no hysteresis at zero magnetic field. In contrast, upon applying a $V_G$ of either +3 V or -3 V, the sample showed clear hysteresis at zero magnetic field. This strongly suggested the occurrence of long-range ordering of magnetic moments in Pt with gating. In addition, the saturation resistance increased with a $V_G$ of either +3 V or -3 V, presumably due to an increase in the magnetism of Pt [17]. The inset shows a coercive field $B_c$ as a function of $V_G$. The $B_c$ was 260 mT and 500 mT for $V_G$ = -3 V and +3 V, respectively. Figure 3(b) shows the $\rho_{xy} - B$ curves for the Pt (7.0 nm)/CFO sample with various $V_G$ values. Without $V_G$, this sample showed a hysteresis loop at zero magnetic field. With $V_G$ of +3 V and -3 V, $B_c$ increased and decreased, respectively, as shown in the inset. The saturation resistance increased for both $V_G$s. Figure 3(c) shows $\rho_{xy} - B$ curves for various $V_G$ values for a sample fabricated on Pt (5.9 nm)/STO. No hysteresis was observed with or without $V_G$. In addition, while an AHE-like curve was observed, the saturation resistance was an order of magnitude smaller than those for the Pt/CFO samples.

Long-range ordering of magnetic moments in Pt atoms was observed in thick Pt films with widths of 7.0 nm by the MPE and was tuned by the electric field effect. However, the electric field effect in the metal is thought to affect Pt atoms as far as 1 nm from the surface due to Thomas-Fermi shielding. Additionally, a previous study [7] showed that the MPE would affect Pt atoms at distances of less than 2 nm from the CFO substrate. Therefore, there are no Pt atoms affected both the MPE and the field effect on our relatively thick Pt films. We speculate that entire Pt atoms in the film has magnetic moments and long-range ordering due to a quantum well state, and so both the MPE and the field-effect affects the long-range ordering. The quantum well state in Pd/STO and Pt/STO films has been reported to enhance the density of states at the Fermi level $D(E_F)$ at a certain thickness, resulting in magnetic moments in the Pd or Pt layer without the MPE [28] [29]. The magnetic moments were periodically changed with changes in thickness with an oscillatory period of around 1 nm. This oscillatory behavior can explain the disappearance and the appearance of the long-range ordering on our 5.9 nm and 7.0 nm Pt films, respectively. Since the Pt (5.9 nm)/CFO film was very near the ferromagnetic state, accumulated charge carriers on the surface would induce long-range ordering with both positive and negative $V_G$s.

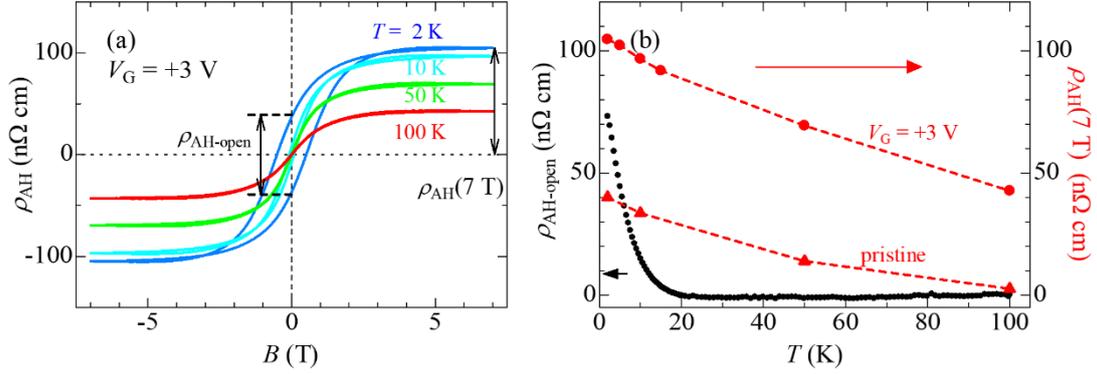

Fig. 4. (a) Magnetic field dependence of $\rho_{AH}$ for Pt (5.9 nm)/CFO with $V_G$ = +3 V, shown for various temperatures (2 K, 10 K, 50 K, and 100 K). (b) Temperature dependence of the resistance opening at 0 T, $\rho_{AH\text{-open}}$, for $V_G$ = +3 V. (left axis, black) Temperature dependence of the saturation value of $\rho_{AH}$ and $\rho_{AH}$(7 T) observed for pristine (marked with Δ) and $V_G$ = +3 V samples (marked with ○) is also shown (right axis, red).

We then measured the temperature dependence of $\rho_{xy}$ for the Pt (5.9 nm)/CFO sample with $V_G$ = +3 V, which was considered the strongest ferromagnetic state. Figure 4(a) shows $\rho_{AH}$ – $B$ curves at various temperatures (2 K, 10 K, 50 K, and 100 K). The $\rho_{OH}$ in Eq. (1) was estimated from the slope over 6 T. Both the magnitude and the width of the hysteresis loop decreased with increasing temperature. Figure 4(b) shows the characteristics of $\rho_{AH}$ measured at each temperature (2 K to 100 K). The hysteresis opening at zero field, $\rho_{AH\text{-open}}$, as observed for typical ferromagnets, was found to exist up to approximately 19.5 K. This indicated that the long-range ordering in the magnetic moments of Pt films disappeared at 19.5 K. In addition, the saturation value of $\rho_{AH}$ ($\rho_{AH}$(7 T)) for $V_G$ = +3 V is always higher than that for $V_G$ = 0 V, suggesting an increase in the saturation value for the magnetic moments of the Pt films.

Finally, we would like to discuss comparisons with other studies. W. Amamou, *et al.* reported large hysteresis, such as perpendicular magnetic anisotropy, at 5 K for Pt (1.7 nm)/CFO/MgO [7]. The effect was not due to spin injection from the CFO because $B_c$ is larger than that for CFO. They concluded that ferromagnetism was induced in Pt by the MPE at the interface because the hysteresis almost disappeared with insertion of the Cu layer. Our CFO/STO sample had in-plane magnetic anisotropy, and $B_c$ was 50 mT with a perpendicular magnetic field. The $B_c$ for the Pt (7.0 nm)/CFO sample was more than 400 mT, which was larger than that of CFO, suggesting that the anomalous Hall resistance was induced by the magnetic moments of Pt rather than by spin injection from CFO, as in W. Amamou, *et al.*

S. Nodo, *et al.* reported a large hysteresis for Pt (2.5 nm)/CFO/MgO that appeared to be perpendicular magnetic anisotropy at 300 K [21]. Because they reported $T_C$ > 300 K in a film identified by XRD as

Pt (111), it is likely that the ferromagnetic ordering in Pt was preserved up to 300 K or more due to appropriate film quality and thickness. In our study, the long-range ordering disappeared at 19.5 K, but a nonlinear AHE was observed up to 250 K. Additionally, we observed a change from a state with no hysteresis to a state with finite $B_c$ depending on the gate bias, whereas S. Nodo, *et al.* showed that $\rho_{AH}$ changed with gate voltage but $B_c$ did not change. Since the sputter-deposited Pt film is probably island-like, slight spatial variations in thickness and crystallinity may cause superparamagnet-like behavior. A ferromagnetically strong part would remain ferromagnetic up to room temperature, but a weak part would be paramagnetic. Assuming that a ferromagnetic interaction of the weak part was enhanced by the gate bias, then the switching of the long-range order by the gate bias can be well explained. This is consistent with the discussion of CW hysteresis observed at large magnetic fields.

In summary, we have successfully controlled the hysteresis loop of $\rho_{AH}$ in Pt/CFO samples by electrolyte gating. A hysteresis loop was observed in the $\rho_{xy} - B$ curve at 2 K for the Pt (7.0 nm)/CFO sample but not for the Pt (5.9 nm)/CFO sample, while both samples exhibited MPE-induced AHE up to 250 K. Application of a $V_G$ of ±3 V to the Pt (5.9 nm)/CFO sample opened the hysteresis loop in the $\rho_{xy} - B$ curve at 2 K, which remained up to 19.5 K. In addition, the magnitude of $\rho_{AH}$ was increased for a $V_G$ of ±3 V. Since the ferromagnetic properties were changed by the electric field effect on a relatively thick (7.0 nm) sample, we speculate that the long-range ordering of the magnetic moments was induced in entire atoms of the Pt film both by the MPE and the quantum well state. Compared with previous reports on Pt/CFO films, we employed a polycrystalline Pt film and a CFO film with in-plane anisotropy. Both weakened the ferromagnetic ordering in Pt and induced superparamagnet-like behavior. Then, the ferromagnetic ordering of the magnetic moments in the Pt film was switched by the electric field effect. If we fabricate Pt films with the appropriate conditions, we can fabricate a room-temperature switching sample of the ferromagnetic hysteresis loop in Pt. Such a sample will be useful for spintronics applications.

This work was supported by JSPS KAKENHI Grant Numbers JP21H01038, JP22K14001 and performed using facilities of the Cryogenic Research Center, the University of Tokyo.

# Supplementary Information

Leon Sakakibara[1], Yumiko Katayama[1], Tomohiro Koyama[2,3,4], Daichi Chiba[2,3,4], and Kazunori Ueno[1]

[1] Graduate School of Arts and Sciences, University of Tokyo, Meguro, Tokyo 153-8902, Japan

[2] Institute of Scientific and Industrial Research, Osaka University, Ibaraki, Osaka 567-0047, Japan

[3] Center for Spintronics Research Network, Osaka University, Toyonaka, Osaka 560-8531, Japan

[4] Spintronics Research Network Division, Institute for Open and Transdisciplinary Research Initiatives, Osaka University, Suita, Osaka 565-0871, Japan

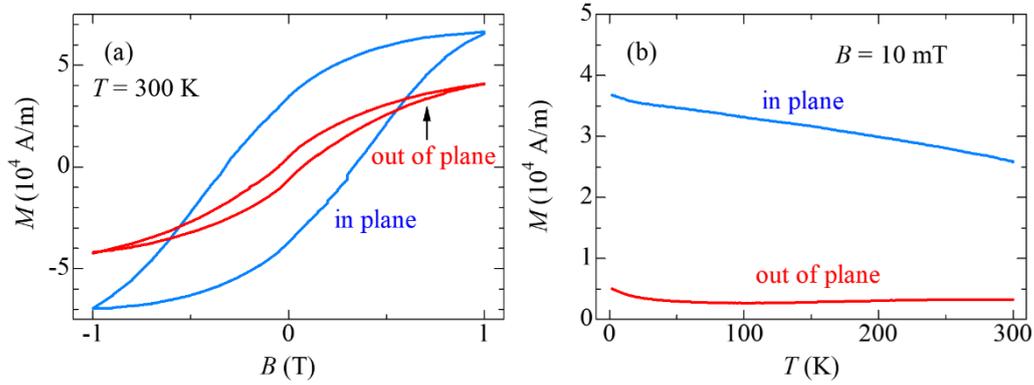

Fig. S1. Magnetization properties of a 50 nm CFO film. (a) Magnetic field dependence at 300 K. (b) Temperature dependence at 10 mT. The magnetic field was applied in the in-plane and out-of-plane directions.

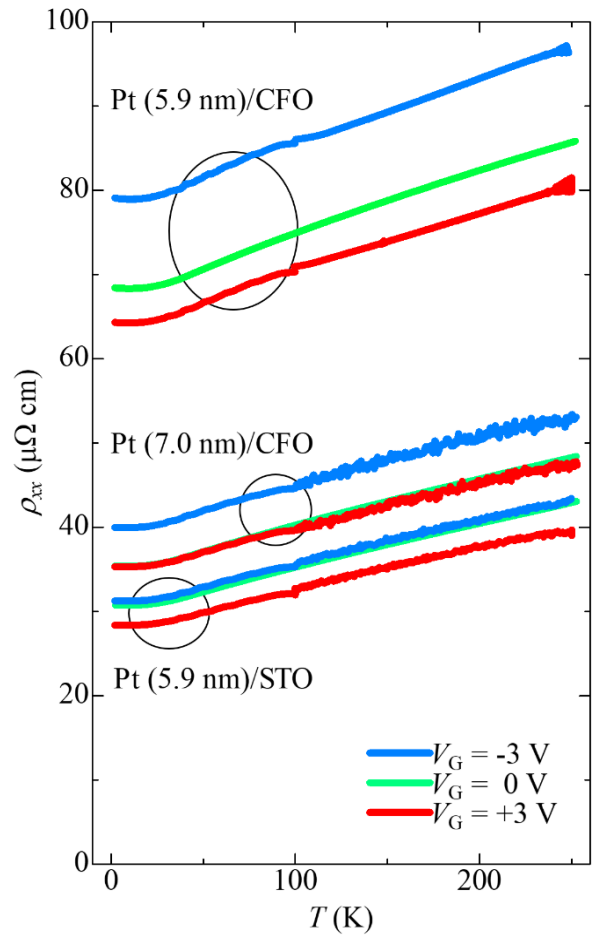

Fig. S2. Temperature dependence of $\rho_{xx}$ for $V_G$ of -3 V (blue), 0 V (green) and +3 V (red) for Pt (5.9 nm)/CFO, Pt (7.0 nm)/CFO and Pt (5.9 nm)/STO samples. Positive and negative gate biases decreased and increased $\rho_{xx}$ for all samples, indicating n-type FET behavior in all samples.

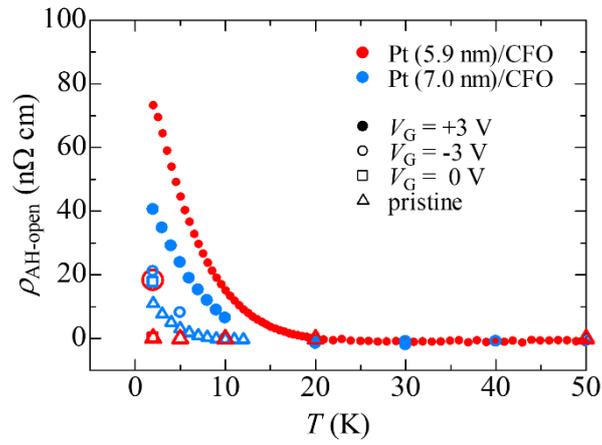

Fig. S3. Temperature dependence of the resistance opening at 0 T, $\rho_{\text{AH-open}}$, for $V_G$ values of +3 V (marked with ●), -3 V (marked with ○), 0 V (marked with □), and pristine (marked with △) for Pt (5.9 nm)/CFO (red) and Pt (7.0 nm)/CFO (blue) samples.